\documentclass
[aps,prd,superscriptaddress,nofootinbib]{revtex4-1}%
\usepackage{amsmath}
\usepackage{amsfonts}
\usepackage{amssymb}
\usepackage{graphicx}
\usepackage{latexsym}
\usepackage{hyperref}
\usepackage{color,epsfig}
\usepackage{ifpdf}
\usepackage{bm}
\usepackage{wasysym}
\usepackage[english]{babel}
\usepackage{braket}
\usepackage{slashed}
\usepackage{longtable,array,cancel}

\hypersetup{colorlinks,citecolor= nicegreen,linkcolor= nicered}
\definecolor{nicered}{rgb}{0.7,0.1,0.1}
\definecolor{nicegreen}{rgb}{0.1,0.5,0.1}

\newcommand{\gs}[1]{\cancel{#1}}

\newcommand{\nn}{\nonumber}
\newcommand{\ii}{\mathrm{i}}
\newcommand{\pr}{\prime}
\newcommand{\bs}[2]{\begin{array}{r}#1\\#2\end{array}}
\newcommand{\LINE}[5]{#1 & #2 & #3 & #4 & #5}

\begin{document}

\title{Probing anomalous $tWb$ interactions with rare $B$ decays}

\author{Jure Drobnak}
\email[Electronic address:]{jure.drobnak@ijs.si}
\affiliation{J. Stefan Institute, Jamova 39, P. O. Box 3000, 1001  Ljubljana, Slovenia} 

\author{Svjetlana Fajfer} 
\email[Electronic address:]{svjetlana.fajfer@ijs.si} 
\affiliation{J. Stefan Institute, Jamova 39, P. O. Box 3000, 1001
  Ljubljana, Slovenia}
\affiliation{Department of Physics,
  University of Ljubljana, Jadranska 19, 1000 Ljubljana, Slovenia}

\author{Jernej F. Kamenik}
\email[Electronic address:]{jernej.kamenik@ijs.si} 
\affiliation{J. Stefan Institute, Jamova 39, P. O. Box 3000, 1001  Ljubljana, Slovenia}
\affiliation{Department of Physics,
  University of Ljubljana, Jadranska 19, 1000 Ljubljana, Slovenia}

\date{\today}

\begin{abstract}
Precision studies of top quark properties are currently underway at the LHC and Tevatron colliders with the prospect of probing anomalous $t-W-b$ interactions.
In the mean time, recent experimental results for the $B_{d,s}- \bar B_{d,s}$ oscillation observables, the branching ratio $B_s \to \mu^+\mu^-$, as well as the forward - backward asymmetry in $B \to K^* \ell^+ \ell^-$, accompanied by the accurate theoretical predictions for the relevant observables obtained within the SM motivate a combined study of these observables in the presence of anomalous $t-W-b$ vertices. We investigate contributions of such anomalous couplings to the $B \to X_s \ell^+ \ell^-$ decay mode, and combining them with the modifications of the $B_{d,s}- \bar B_{d,s}$, $ B\to X_s \gamma$ observables, we determine indirect bounds on the real and imaginary parts of the anomalous $t-W-b$ interactions. We find these to be mostly superior to present direct constraints coming from top decay and production measurements at the LHC and Tevatron. Finally, we  predict the allowed effects in the branching ratios of the $B_s \to \mu^+ \mu^-$ and $B \to K^{(*)} \nu \bar \nu$, as well as the forward-backward asymmetry in $B \to K^* \ell^+ \ell^-$. We find that improved knowledge of these observables in the future could further constrain some of the anomalous $tWb$ interactions.
\end{abstract}

\maketitle

\section{Introduction}

In many scenarios of physics beyond the Standard Model (SM), the top quark provides a preferred search window due to its large coupling to the physics responsible for the electro-weak symmetry breaking. 
A fascinating possibility then is that the top quark properties exhibit deviations from their predictions within the SM. The leading such contributions are encoded in the anomalous top couplings to SM gauge fields affecting the top anomalous (chromo)-electric and (chromo)-magnetic dipole moments~\cite{Kamenik:2011dk}, Flavor Changing Neutral Current (FCNC) couplings ($tVu_j$, where $V=g,\gamma,Z$)~\cite{Fox:2007in}, as well as anomalous charged current interactions ($tWd_j$).

Precision studies of top quark properties are currently underway at the LHC and Tevatron colliders and will complete our knowledge of possible New Physics (NP) effecting the properties and interactions of the SM particles. In particular,  both Tevatron~\cite{oldHelicity,newHelicity} and LHC~\cite{ATLAS-CONF-2011-122} experiments have recently presented precise measurements, sensitive to the top quark -- $W$ boson -- $b$ quark  vertex ($tWb$). 

Recently, we have reconsidered the $t \to b W$ helicity amplitudes in presence of anomalous $tWb$ interactions, including corrections due to perturbative QCD effects~\cite{Drobnak:2010ej}, and found that some of the anomalous couplings are already being significantly constrained by the CDF and D\O\, measurements. In light of the first LHC results being able to provide constraints on possible anomalous $tWb$ couplings,  a more recent study~\cite{AguilarSaavedra:2011ct} has combined bounds coming from  $t \to b W$ helicity amplitudes, angular asymmetries as well as single top production observables.

On the other hand, the $tWb$ vertex also plays an important role in $B$ physics, since FCNC processes involving $b$ quarks receive, within the SM, dominant contributions from loops involving a top quark and a $W$ boson. During the recent years, the $B$ factories, CDF and D\O\, experiments at the Tevatron as well as LHCb at the LHC have proceeded to probe $B_{d,s}- \bar B_{d,s}$ oscillation observables,  as  well as rare FCNC mediated $B$ decays with ever increasing precision. 
The first $B$ physics observable to be studied for the effects of the anomalous $tWb$ vertices was the $B \to X_s \gamma$ decay rate~\cite{Grzadkowski:2008mf}. The authors considered a complete basis of dimension six operators, free of tree-level FCNCs. To simplify the analysis, they assumed alignment of the operator flavor structures with the SM CKM matrix, as well as the reality of the appropriate Wilson coefficients. 
Recently~\cite{Drobnak:2011wj}, we have explored the possible presence of  anomalous $tWd_j$ interactions  in $B_{d,s}- \bar B_{d,s}$ oscillations within a model independent approach based on the assumptions of Minimal Flavor Violation (MFV)~\cite{D'Ambrosio:2002ex}, which extends the basis of operator flavor structures considered in~\cite{Grzadkowski:2008mf}. We have found that the $\Delta B=2$ mixing amplitudes can receive large contributions and that some of the possible operators can provide significant new sources of CP violation.  In term, based on the recent global fits to neutral meson oscillation observables~\cite{Lenz:2010gu}, we have derived preferred ranges for the anomalous couplings, some of them being more constraining than both $B \to X_s \gamma$ as well as direct measurements.

Inspired by the fast progress in the field we extend our analysis of the anomalous $tWd_j$ vertices to rare (semi)leptonic $B$ decays. Again we rely on MFV in order to reduce the number of possible operator flavor structures, and we review the framework in Sec. II.
After performing the one-loop matching of our operator basis onto the low energy effective Hamiltonian describing $\Delta B=1$ FCNC processes, we obtain corrections to the relevant Wilson coefficients in Sec. III. 
We proceed in Sec. IV by calculating the effects in the inclusive $B\to X_s\gamma$ and $B \to  X_s \ell^+ \ell^-$ decays, cross-checking the existing calculation~\cite{Grzadkowski:2008mf} in the former mode.
In order to derive bounds on both the real and imaginary parts of the appropriate Wilson coefficients we include the experimental results not only for the decay rates but also for the CP asymmetry in $B \to X_s \gamma$. 
After performing a global fit of the Wilson coefficients, we derive predictions for several rare $B$ meson processes like $B_s \to \mu^+ \mu^-$, the forward-backward asymmetry (FBA) in $B \to K^* \ell^+ \ell^-$ and the  branching ratios for $B \to K^{(*)} \nu \bar \nu$ in Sec. V.  Finally we conclude in Sec. VI.


\section{Framework}
We adopt the framework used in our previous work~\cite{Drobnak:2011wj}, committing to an effective theory, described by the Lagrangian
\begin{eqnarray}
{\cal L}={\cal L}_{\mathrm{SM}}+\frac{1}{\Lambda^2}\sum_i C_i \mathcal Q_i +\mathrm{h.c.}+ {\cal O}(1/\Lambda^3)\,,
\label{eq:lagr}
\end{eqnarray}
where ${\cal L}_{\mathrm{SM}}$ is the SM part, $\Lambda$ is the scale of NP and ${\cal Q}_i$ are dimension-six operators, invariant under SM gauge transformations and consisting of SM fields. We assume that at the scale $\mu\sim m_t$ the SM fields with up to two Higgs doublets are the only propagating degrees of freedom, that the electroweak symmetry is only broken by the vacuum expectation values of these two scalars and that operators up to dimension six give the most relevant contributions to the observables we consider. Such an approach is appropriate to summarize weak scale effects of NP at $\Lambda\gg m_t$, where the new heavy degrees of freedom have been integrated out.

Our operator basis consists of all dimension-six operators that generate charged current quark interactions with the $W$, but do not induce FCNCs at the tree-level. Since we restrict our discussion to MFV scenarios, Lagrangian (\ref{eq:lagr}) has to be formally invariant under the SM flavor group $\mathcal G^{\rm SM} = U(3)_{Q} \times U(3)_u \times U(3)_d$ where $Q, u, d$ stand for quark doublets and up and down type quark singlets respectively. MFV requires that the only $\mathcal G^{\rm SM}$ symmetry breaking spurionic fields in the theory are the up and down quark Yukawa matrices $Y_{u,d}$, formally transforming under $\mathcal G^{\rm SM}$ as $(3,\bar 3,1)$ and $(3,1,\bar 3)$ respectively. 

We identify four relevant quark bilinears with distinct transformation properties under $\mathcal G^{\rm SM}$: $\bar u d$, $\bar Q Q$, $\bar Q u$ and $\bar Q d$ transforming as $(1,\bar 3, 3)$, $({1\oplus8},1,1)$, $(\bar 3,3,1)$ and $(\bar 3, 1, 3)$ respectively. Using these, we can construct the most general $\mathcal G^{\rm SM}$ invariant quark bilinear flavor structures as
\begin{equation}
\bar u Y_u^\dagger \mathcal A_{ud} Y_d d\,, ~~~ \bar Q \mathcal A_{QQ} Q\,, ~~~ \bar Q \mathcal A_{Qu} Y_u u\,,~~~ \bar Q \mathcal A_{Qd} Y_d d\,,
\label{eq:flav}
\end{equation}
where $\mathcal A_{xy}$ are arbitrary polynomials of $Y_{u} Y_{u}^\dagger$ and/or $Y_{d}Y_d^\dagger$,  transforming as $({1\oplus8},1,1)$.

In order to identify the relevant flavor structures in terms of physical parameters, we can without the loss of generality consider $Y_{u,d}$ condensate values in a basis in which $\langle Y_d \rangle$ is diagonal: $\langle Y_d \rangle =\mathrm{diag}(m_d,m_s,m_b)/v_d$ and $\langle Y_u \rangle =V_{}^\dagger \mathrm{diag}(m_u,m_c,m_t)/v_u$, where we have introduced separate up- and down-type Higgs condensates $v_{u,d}$, while $V_{}$ is the SM CKM matrix. We also write $Q,u,d$ in this basis in terms of  quark mass eigenstates $u_{Li}, d_{Li}, u_{Ri}, d_{Ri}$,
where $L,R$ subscripts denote chirality projectors $\psi_{R,L} = (1\pm \gamma_5) \psi/2$.

The final specification of the operator basis has been attentively described in Ref.~\cite{Drobnak:2011wj}. It consists of seven distinct operators
\begin{eqnarray}
\nn \mathcal Q_{LL}&=&[\bar Q^{\prime}_3\tau^a\gamma^{\mu}Q'_3] \big(\phi_d^\dagger\tau^a\ii D_{\mu}\phi_d\big) \hspace{-0.1cm}-\hspace{-0.1cm}[\bar Q'_3\gamma^{\mu}Q'_3]\big(\phi_d^\dagger\ii D_{\mu}\phi_d\big),\\
\nn \mathcal Q'_{LL}&=&[\bar Q_3\tau^a\gamma^{\mu}Q_3] \big(\phi_d^\dagger\tau^a\ii D_{\mu}\phi_d\big) \hspace{-0.1cm}-\hspace{-0.1cm}[\bar Q_3\gamma^{\mu}Q_3]\big(\phi_d^\dagger\ii D_{\mu}\phi_d\big),\\
\nn \mathcal Q^{\prime\prime}_{LL}&=&V_{tb}^{*}\Big\{[\bar Q'_3\tau^a\gamma^{\mu}Q_3] \big(\phi_d^\dagger\tau^a\ii D_{\mu}\phi_d\big) \hspace{-0.1cm}
-[\bar Q'_3\gamma^{\mu}Q_3]\big(\phi_d^\dagger\ii D_{\mu}\phi_d\big)\Big\}\,,\\
\nn \mathcal Q_{RR}&=& V_{tb} [\bar{t}_R\gamma^{\mu}b_R] \big(\phi_u^\dagger\ii D_{\mu}\phi_d\big) \,, \nn\\
\nn \mathcal Q_{LRb} &=& [\bar Q_3 \sigma^{\mu\nu}\tau^a b_R]\phi_d W_{\mu\nu}^a \,,\\
\nn \mathcal Q_{LRt} &=& [\bar Q'_3 \sigma^{\mu\nu}\tau^a t_R]{\phi_u}W_{\mu\nu}^a \,,\\
\mathcal Q'_{LRt} &=&V_{tb}^{*} [\bar Q_3 \sigma^{\mu\nu}\tau^a t_R]{\phi_u}W_{\mu\nu}^a \,,
\label{eq:ops1}
\end{eqnarray}  
where we have introduced the left-handed $SU(2)$ doublets
\begin{eqnarray}
Q_3=(V^*_{ib} u_{Li},b_{L})\,,\hspace{0.3cm} Q'_3 = (t_L,V_{tj}d_{jL})\,,
\end{eqnarray}
with $i,j$ flavor indices. Furthermore we have
\begin{eqnarray}
D_{\mu}&=&\partial_{\mu}+\ii \frac{g}{2}A_{\mu}^a\tau^a +\ii \frac{g'}{2}B_{\mu} Y\,, \nonumber \\
W^a_{\mu\nu}&=&\partial_{\mu}W_{\nu}^a-\partial_{\nu}W_{\mu}^a - g\epsilon_{abc}W_{\mu}^b W_{\nu}^c\,,
\end{eqnarray}
and finally $\sigma^{\mu\nu}=\ii [\gamma^{\mu},\gamma^{\nu}]/2$, while $\phi_{u,d}$ are the up- and down-type Higgs fields (in the SM $\phi_u =\ii \tau^2 \phi_d^*$). As already mentioned in the introduction, our operator basis coincides with the one used in Ref.~\cite{Grzadkowski:2008mf}, expanded by the three primed operators. Furthermore we do not make the operators hermitian, hence effects of operators ${\cal Q}_{i}^{\dagger}$ are accompanied by $C_i^{*}$ and will be kept track of separately. For completeness we note that in scenarios, where the $Y_d \, (\sim m_b/v_d)$ expansion is perturbative, the operators $\mathcal Q_{LL}$ and $\mathcal Q_{LRt}$ are expected to dominate, with $\mathcal Q_{RR}$ and $\mathcal Q_{LRb}$ being suppressed by a single power of $Y_d$ and the operators $\mathcal Q^{\prime(\prime)}_{LL}$ and $\mathcal Q'_{LRt}$ being supressed by $Y_d Y_d^\dagger$.


\section{Matching}
To establish how operators (\ref{eq:ops1}) affect $b\to s \gamma,g$ and $b\to s \ell^+ \ell^-,\nu\bar{\nu}$ transitions, we match our effective theory (\ref{eq:lagr}), to the low energy effective theory described by
\begin{eqnarray}
{\cal L}_{\mathrm{eff}}&=& {\cal L}^{}_{\mathrm{QCD\times QED}}   + \frac{4 G_F}{\sqrt{2}}\Big[ \sum_{i=1}^2 C_{i}( V_{ub}V_{us}^*\mathcal O^{(u)}_i +  V_{cb}V_{cs}^* \mathcal O^{(c)}_i) \Big] + \frac{4 G_F}{\sqrt{2}}V_{tb}V_{ts}^*\Big[\sum_{i=3}^{10} C_{i}{\cal O}_i +  C_{\nu\bar{\nu}}{\cal O}_{\nu\bar{\nu}}\Big]\,, \label{eq:loweff}
\end{eqnarray}
where the first term consists of kinetic terms of the light SM particles as well as their QCD and QED interactions.  The relevant operators read
\begin{align}
{\cal O}_7&= \frac{e m_b}{(4\pi)^2}\big(s_L\sigma_{\mu\nu}b_R\big)F^{\mu\nu}\,,&
{\cal O}_9&= \frac{e^2}{(4\pi)^2}\big(s_L\gamma^{\mu}b_L\big)\big(\bar{\ell}\gamma_{\mu}\ell\big)\,, \nonumber \\
{\cal O}_8&= \frac{g_s m_b}{(4\pi)^2}\big(s_L\sigma_{\mu\nu}T^ab_R\big)G_a^{\mu\nu}\,,&
{\cal O}_{10}&= \frac{e^2}{(4\pi)^2}\big(s_L\gamma^{\mu}b_L\big)\big(\bar{\ell}\gamma_{\mu}\gamma_5\ell\big)\,,\nonumber \\
{\cal O}_{\nu\bar{\nu}}&=\frac{e^2}{(4\pi)^2}\big(\bar{s}_L\gamma^{\mu}b_L\big)\big(\bar{\nu}\gamma_{\mu}(1-\gamma^5)\nu\big) \,. \label{eq:ops2}
\end{align}
On the other hand, since they are not that crucial for our analysis, we omit the definition of the four-quark operators ${\cal O}_{1,\dots,6}$ which can be found for example in Ref.~\cite{Buchalla:1995vs}.

\begin{figure}[h]
\begin{center}
\includegraphics[scale= 0.6]{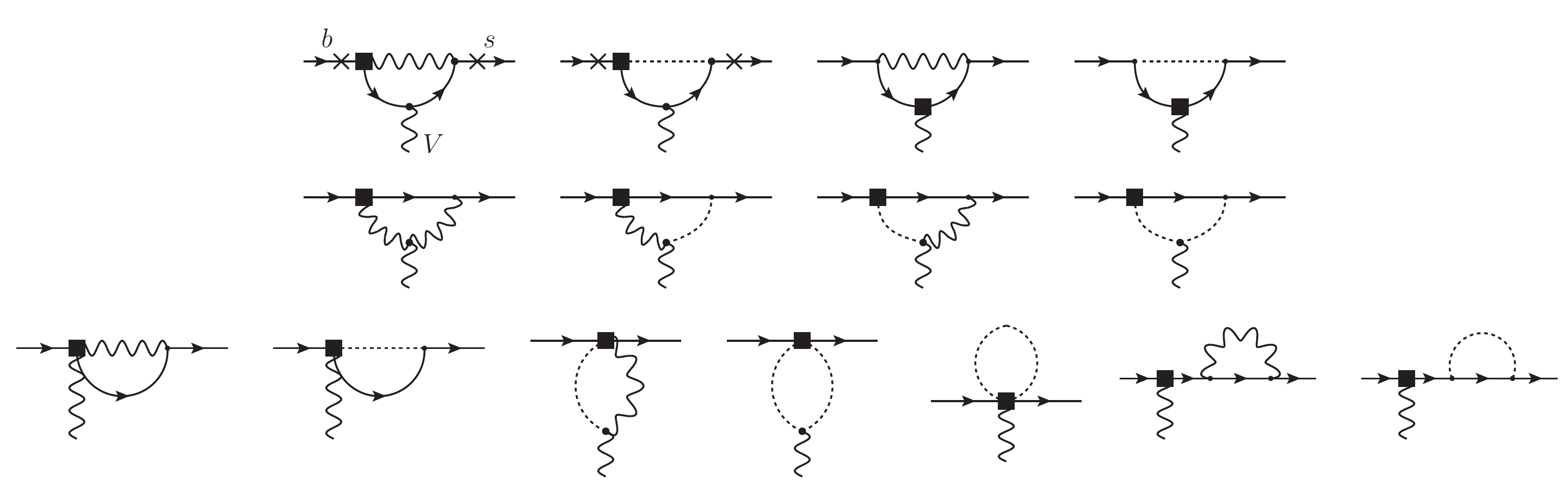}
\caption{Types of diagrams encountered when computing $b\to s V$ transitions, where $V$ stands for $\gamma, Z, g$. Dotted lines represent would-be Goldstone bosons, crosses mark additional points where $V$ can be emitted in one-particle-reducible diagrams and square represents an anomalous coupling. Gluon emission is only possible from quark lines and with the SM coupling. Quarks running in the loops are of up-type.}
\label{fig:feyns11}
\end{center}
\end{figure}

\begin{figure}[h]
\begin{center}
\includegraphics[scale= 0.6]{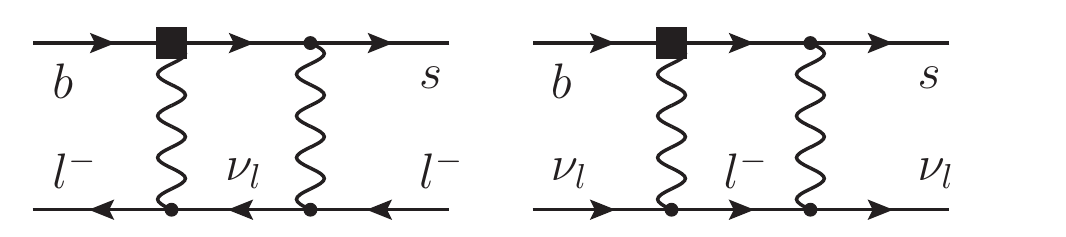}
\caption{Box diagrams contributing to $b\to s \ell^+ \ell^-$ and $b\to s \nu\bar{\nu}$ transitions. Diagrams with would-be Goldstons are absent, since the leptons are treated as massless.}
\label{fig:feyns3}
\end{center}
\end{figure}

We perform the matching of the operators in \eqref{eq:lagr} to \eqref{eq:loweff} by integrating out the top quark and  electroweak bosons at leading order (LO) in QCD. 
We perform our calculation in a general $R_{\xi}$ gauge for the weak interactions, giving us the opportunity to check the $\xi$ cancelation in the final results. The drawback is the appearance of would-be Goldstone bosons in both SM as well NP contributions. As a consequence of formulating the effective theory in a gauge and MFV invariant manner, our operators contribute to processes of interest with grater intricacy than just a mere modification of the $tWb$ vertex. Generic penguin and box diagrams with anomalous couplings are shown in Fig.~\ref{fig:feyns11} and Fig.~\ref{fig:feyns3}. Exact diagrams for a specific ${\cal Q}_i$ can be reconstructed using Feynman rules given in the Appendix~\ref{sec:app1}.

In all calculations we neglect the $s$ quark and lepton masses. When dealing with $b\to s\gamma,g$ transitions, we expand the amplitudes up to second order in external momenta and keep first order ${\cal O}(m_b)$ terms.  This allows us to generate operators ${\cal O}_{7,8}$ as well as photonic contributions to ${\cal O}_9$. On the other hand, in the effective $b\to s Z$  and box diagram contributions presented in Fig.~\ref{fig:feyns3}, which both contribute to ${\cal O}_{9,10,\nu\bar{\nu}}$, the external momenta can be completely neglected. We are interested in NP contributions to the observables at order $1/\Lambda^2$ and thus only need to consider single operator insertions. 


As the result of the matching procedure we obtain deviations from the SM values for the Wilson coefficients, i.e. $C_i = C_i^{\rm SM} + \delta C_i$, which we parametrize as
\begin{eqnarray}
\delta C_i(\mu) &=&\sum_{j}\kappa_j(\mu) f_i^{(j)}(x_t,\mu) + \kappa_j^{*}(\mu) \tilde{f}^{(j)}_i (x_t,\mu)\,,\label{eq:fs}
\end{eqnarray}
where $x_t=\overline{m}_t^2/m_W^2$, $\mu$ is the matching scale and $\kappa_j$ are defined as
\begin{eqnarray}
\kappa_{LL}^{(\prime,\prime\prime)}=\frac{C_{LL}^{(\prime,\prime\prime)}}{\Lambda^2\sqrt{2}G_F}\,,\hspace{0.3cm}
\kappa_{RR}=\frac{C_{RR}}{\Lambda^2 2\sqrt{2} G_F}\,,\hspace{0.3cm}
\kappa_{LRb}=\frac{C_{LRb}}{\Lambda^2 G_F}\,,\hspace{0.3cm}
\kappa_{LRt}^{(\prime)}=\frac{C_{LRt}^{(\prime)}}{\Lambda^2 G_F}\,.
\label{eq:kappas}
\end{eqnarray}
Analytical expressions for $f_i^{(j)}, \tilde{f}_i^{(j)}$ are given in the Appendix~\ref{sec:app2}. 

We note, that the matching procedure for operator ${\cal Q}_{LL}^{\pr}$ stands out compared to the other operators. The charged current structure of this operator resembles that of the SM operator $\bar{Q}_i\gamma^{\mu}\tau^a Q_i W^a_\mu$. Consequently Wilson coefficients $C_{1,\dots,8}$ are changed in a trivial way $C_i(\mu) = (1+\kappa_{LL}^{\pr}(\mu))C_i^{\mathrm{SM}}(\mu)$. The change of the remaining Wilson coefficients $C_{9,10,\nu\bar{\nu}}$, matching of which involves the $Z$ boson, is however not of this form.

Some of the diagrams in Fig.~\ref{fig:feyns11} are UV divergent. We remove these divergences using the $\overline{\mathrm{MS}}$ prescription leading to remnant $\log (m_W^2/\mu^2)$ terms. We shall quantify the matching scale dependence of our results and consequently their sensitivity to the UV completion of the effective theory by varying the scale between $\mu=2m_W$ and $\mu=m_W$.
Since UV renormalization is necessary, our operator basis indeed needs to be extended to include operators that can serve as the appropriate counter-terms. Within the employed MFV framework examples of these operators read
\begin{eqnarray}
{\cal Q}_1=\big[\bar{Q}\sigma^{\mu\nu}A_{Qd} Y_d\tau^a d\big]\phi_d W^a_{\mu\nu}\,,\hspace{0.5cm}
{\cal Q}_2=\big[\bar{Q}\gamma^{\mu}A_{QQ}Q\big]\big[\bar{\ell}\gamma_{\mu}\ell\big]\,, \nonumber\\
{\cal Q}_3=\big[\bar{Q}\gamma^{\mu}\tau^a A^{\pr}_{QQ}Q\big]\big[\phi_d^{\dagger}\tau^a\ii D_{\mu}\phi_d\big] + \big[\bar{Q}\gamma^{\mu} A^{\pr}_{QQ}Q\big]\big[\phi_d^{\dagger}\ii D_{\mu}\phi_d\big]\,.
\label{eq:counterterms}
\end{eqnarray}
The operator ${\cal Q}_1$ produces a counter-term for divergences in $\delta C_7$ , while ${\cal Q}_{2,3}$ provide counter-terms for divergent parts of $\delta C_{9,10,\nu\bar{\nu}}$. The operator ${\cal Q}_3$ generates a tree-level $bZs$ vertex. 
The sets of flavor  matrices needed to match the structures of divergencies generated by the various operators in \eqref{eq:ops1} are
\begin{eqnarray}
A_{Qd}&=&Y_u Y_u^{\dagger}\,,\nonumber \\
A_{QQ}&=&Y_u Y_u^{\dagger}\,,\, Y_u Y_u^{\dagger} Y_d Y_d^{\dagger}\,,\nonumber \\
A_{QQ}^{\pr}&=&Y_u Y_u^{\dagger}\,,\, (Y_u Y_u^{\dagger})^2\,,\,Y_u Y_u^{\dagger}Y_d Y_d^{\dagger}\,,\, (Y_u Y_u^{\dagger})^2Y_d Y_d^{\dagger}\,,\,
Y_u Y_u^{\dagger} Y_d Y_d^{\dagger}Y_u Y_u^{\dagger}\,.
\end{eqnarray}
In our numerical analysis, we will selectively set contributions of certain operators to be nonzero. Consequently in the following, we will drop the implicit (tree-level) contributions of the operators in~\eqref{eq:counterterms} to $\delta C_i$, as these have been already investigated and constrained in the existing literature~\cite{Hurth:2008jc}.

All $f_i^{(j)}, \tilde{f}_i^{(j)}$ are found to be $\xi$ independent and a crosscheck with results from Ref.~\cite{Grzadkowski:2008mf} is possible for some of them. We confirm their original results for all the operators except $\mathcal Q_{LRb}$, while an updated version of~\cite{Grzadkowski:2008mf} confirms our result also for this operator. 

\begin{table}[h]
\begin{tabular}{c||c|cc|cc|cc|cc|cc|cc}
&SM&$\kappa_{LL}$&$\kappa_{LL}^*$&$\kappa_{LL}^{\pr}$&$\kappa_{LL}^{\pr*}$&$\kappa_{LL}^{\pr\pr}$&$\kappa_{LL}^{\pr\pr*}$&$\kappa_{RR}$&$\kappa_{LRb}$&$\kappa_{LRt}$&$\kappa_{LRt}^*$&$\kappa_{LRt}^{\pr}$&$\kappa_{LRt}^{\pr*}$\\\hline\hline
$f_7$&-0.19 &0.45& 0.45& -0.19& 0& 0.45& 0& -45.3& 85.5& -0.13& -0.17& -0.15& -0.17\\
$f_8$&-0.095&0.24& 0.24& -0.095& 0& 0.48& 0& -20.2& 54.5& 0.15& 0.05& 0& 0.05\\
$f_9$&1.34&-1.11& -1.11& 1.35& 0.09& -1.11& 0.009& 0& 0& 0.64& 0.64& 0.009& 0.64\\
$f_{10}$&-4.16&1.48& 1.48& -4.28& -0.12& 1.48& -0.12& 0& 0& -2.41& -2.41& 0& -2.41 \\
$f_{\nu\bar{\nu}}$&-6.52&2.38& 2.38& -6.63& -0.12& 2.38& -0.12& 0& 0& -4.25& -4.25& 0& -4.25\\
\end{tabular}
\caption{Numerical values of functions $f_i^{(j)}$ and $\tilde{f}_i^{(j)}$ at $\mu=2 m_W$. Numerical values used for the input parameters are $\overline{m}_t(2m_W)=165.0$~GeV, $s_W^2=0.231$,  $m_W= 80.4$~GeV, $\overline{m}_b(2m_W)= 2.9$~GeV, $|V_{tb}|^2=1$. All $f_i$ values correspond to matching at LO in QCD.}
\label{tab:cs}
\end{table}

To quantify the effects of our seven operators on Wilson coefficients (\ref{eq:loweff}) we present the numerical values of $f_i^{(j)}$ evaluated at $\mu=2m_W$ in Tab.~\ref{tab:cs}. We see that the contributions of the operator ${\cal Q}_{LL}$ and ${\cal Q}_{LL}^{\dagger}$ are identical in all cases, which means that $\kappa_{LL}$ can not induce new CP violating phases in the Wilson coefficients. Likewise, $\mathcal Q_{LRt}$ contributions to $C_{9,10,\nu\bar\nu}$ are Hermitian. On the other hand it can induce a new CP violating phase in $C_{7,8}$. 
Finally at order $1/\Lambda^2$, operators ${\cal Q}_{RR}$ and ${\cal Q}_{LRb}$ which contain right-handed down quarks only contribute to $C_{7,8}$. These contributions are however very significant, since they appear enhanced as $m_t/m_b$ (\ref{eq:rr}, \ref{eq:lrb}) due to the lifting of the chiral suppression, as already pointed out in Ref.~\cite{Grzadkowski:2008mf}.


\section{Bounds on anomalous couplings}\label{sec:bounds}
Having computed $\delta C_i$ in terms of $\kappa_{i}$, we turn our attention to observables affected by such contributions. In particular at order $1/\Lambda^2$, the presently most constraining observables -- the decay rates for $B\to X_s\gamma$ and $B\to X_s \ell^+\ell^-$  are mostly sensitive to the real parts of $\kappa_i$~\cite{Huber:2005ig}.  While in general both $B\to X_{d,s} \gamma$ channels are complementary in their sensitivity to flavor violating NP contributions~\cite{Crivellin:2011ba},  within MFV such effects are to a very good approximation universal and the smaller theoretical and experimental uncertainties in the later mode make it favorable for our analysis. In order to bound imaginary parts of $\kappa_i$, we consider the CP asymmetry in $B\to X_s \gamma$. Finally, we compare and combine these bounds with the ones obtained from $B_{d,s} - \bar B_{d,s}$ oscillation observables in~\cite{Drobnak:2011wj}. 

\subsection{Real parts}
We consider the inclusive $B\to X_s\gamma$ and $B\to X_s \ell^+\ell^-$ branching ratios, for which the presently most precise experimental values have been compiled in~\cite{Asner:2010qj, Huber:2007vv}
\begin{align}
&\mathrm{Br}[\bar{B}\to X_s \gamma]_{E_{\gamma}>1.6~\mathrm{GeV}}=(3.55 \pm 0.26)\times 10^{-4}\,, \nonumber \\
&\mathrm{Br}[\bar{B}\to X_s \mu^+ \mu^-]_{\mathrm{low}\, q^2}\hspace{0.2cm}=(1.60\pm 0.50)\times 10^{-6}\,.
\end{align}

Because the SM contributions to $C_{i}(\mu_b)$ and the corresponding operator matrix elements are mostly real~\cite{Huber:2005ig}, the linear terms in $\delta C_i$, which stem from SM--NP interference contributions contribute mostly as $\mathrm{Re}[\delta C_i]$. 
These are the only terms contributing at order $1/\Lambda^2$.
Therefore, the bounds derived from these two observables are mostly sensitive to the real parts of $\kappa_j$.  Using results of~\cite{Huber:2005ig}, we have explicitly verified that the small $\mathrm{Im}[\delta C_i]$ contributions to ${\mathrm{Br}} [\bar B\to X_s \ell^+\ell^-]$ have negligible effect for all operators except $\mathcal Q_{RR,LRb}$. However, even for these operators ${\rm Im}[\kappa_i]$ are much more severely constrained by $\mathcal A_{X_s \gamma}$, discussed in the next section. Also, using known NLO $B\to X_s\gamma$ formulae~\cite{Chetyrkin:1996vx}, we have verified that ${\rm Im}[\delta C_i]$ contributions to this decay rate are negligible. To analyze the effects of $\delta C_i$ on the two branching ratios, we therefore neglect the small $\mathrm{Im}[\delta C_i]$ contributions and employ the semi-numerical formulae given in Ref.~\cite{DescotesGenon:2011yn} with a few modifications that we specify below:
\begin{itemize}
\item In~\cite{DescotesGenon:2011yn} all predictions are given in terms of $\delta C_i$ at the $b-$scale $\mu_b=4.8$~GeV. Since we wish to check how our results depend on the matching scale $\mu$, we express $\delta C_i(\mu_b)$ using NNLO QCD running~\cite{NNLORGE} as
\begin{eqnarray}
\delta C_7(\mu_b) &=& 0.627\,\delta C_7(m_W)\,,\nonumber\\
\delta C_7(\mu_b) &=& 0.579\,\delta C_7(2m_W)\,.
\end{eqnarray}
On the other hand, $C_{9,10}$ are only affected by EW running and their change with scale from $2m_W$ to $m_W$ is negligible. 
\item The authors of Ref.~\cite{DescotesGenon:2011yn} assumed $\delta C_8 =0$, which is not the case in our analysis. However, LO $C_7$ and $C_8$ ( thus also $\delta C_7$ and $\delta C_8$) enter both observables in approximately the same combination (conventionally denoted as $C_7^{eff}$, c.f.~\cite{Buras:1993xp}). Employing the known SM NNLO matching and RGE running formulae~\cite{NNLORGE} we can correct for this with a simple substitution in the expressions of Ref.~\cite{DescotesGenon:2011yn} for the branching ratios
$\delta C_7 \to \delta C_7 + 0.24\, \delta C_8$, where we have neglected the small difference between the matching conditions at $\mu = 2m_W$ and $\mu=m_W$. We have verified that in this way we reproduce approximately the known $\delta C_8$ dependencies in $B\to X_s\gamma$~\cite{Freitas:2008vh} and $B\to X_s \ell^+\ell^-$~\cite{Huber:2005ig}. 
\item As pointed out in the previous section, ${\cal Q}_{LL}^{\pr}$ is to be treated differently than the other operators. Its effects in $\mathcal O_{3,\dots,8}$ can be seen as a shift in the CKM factor appearing in Eq.~(\ref{eq:loweff}) $V_{tb}V_{ts}^*\to(1+\kappa_{LL}^{\pr})V_{tb}V_{ts}^*$. Consequently SM predictions for these contributions simply get multiplied by the factor of $|1+\kappa_{LL}^{\pr}|^2$ and only $\delta C_{9,10,\nu\bar{\nu}}$ need to be considered as nonzero.
\end{itemize}
Taking all this into account and considering only one operator ${\cal Q}_i$ to contribute at a time, we obtain the 95\% C.L. bounds on ${\rm Re} [\kappa_i]$ shown in Tab.~\ref{tab:bounds}.
\begin{table}[h!]
\hspace{-1cm}
\begin{minipage}{0.55 \textwidth}
\begin{center}
\begin{tabular}{c||ccc|c}
&$B-\bar{B}$&$B\to X_s\gamma$&$B\to X_s \mu^{+}\mu^-$ & combined \\
\hline\hline
\LINE{$\kappa_{LL}$}{$\bs{0.08}{-0.09}$}
{$\bs{0.03}{-0.12}$}
{$\bs{0.48}{-0.49}$}
{$\bs{0.04}{-0.09}\Big(\bs{0.03}{-0.10}\Big)$}\\\hline
\LINE{$\kappa_{LL}^{\pr}$}{$\bs{0.11}{-0.11}$}
{$\bs{0.17}{-0.04}$}
{$\bs{0.31}{-0.30}$}
{$\bs{0.11}{-0.06}\Big(\bs{0.10}{-0.06}\Big)$}\\\hline
\LINE{$\kappa_{LL}^{\pr\pr}$}{$\bs{0.18}{-0.18}$}
{$\bs{0.06}{-0.22}$}
{$\bs{1.02}{-1.04}$}
{$\bs{0.08}{-0.17}\Big(\bs{0.05}{-0.15}\Big)$}\\\hline
\LINE{$\kappa_{RR}$}{}
{$\bs{0.003}{-0.0006}$}
{$\bs{0.68}{-0.66}^*$}
{$\bs{0.003}{-0.0006}\Big(\bs{0.002}{-0.0006}\Big)$}\\\hline
\LINE{$\kappa_{LRb}$}{}
{$\bs{0.0003}{-0.001}$}
{$\bs{0.34}{-0.35}^*$}
{$\bs{0.0003}{-0.001}\Big(\bs{0.003}{-0.01}\Big)$}\\\hline
\LINE{$\kappa_{LRt}$}{$\bs{0.13}{-0.14}$}
{$\bs{0.51}{-0.13}$}
{$\bs{0.38}{-0.37}$}
{$\bs{0.13}{-0.07}\Big(\bs{0.12}{-0.14}\Big)$}\\\hline
\LINE{$\kappa_{LRt}^{\pr}$}{$\bs{0.29}{-0.29}$}
{$\bs{0.41}{-0.11}$}
{$\bs{0.75}{-0.73}$}
{$\bs{0.27}{-0.07}\Big(\bs{0.25}{-0.06}\Big)$}\\\hline
\end{tabular}
\end{center}
\end{minipage}
\hspace{0.2cm}
\begin{minipage}{0.4\textwidth}
\begin{center}
\includegraphics[scale= 0.7]{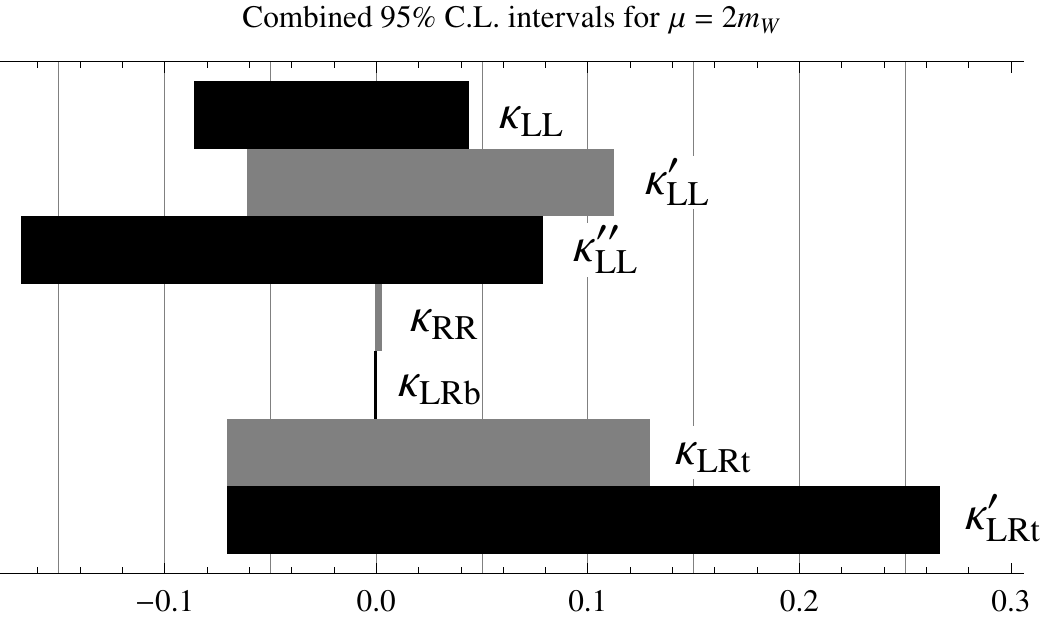}
\end{center}
\end{minipage}
\caption{Lower and upper $95\%$ C.L. bounds on real parts of individual anomalous couplings $\kappa_j$ for $\mu = 2 m_W$ and bracketed for $\mu=m_W$, where $\overline{m}_t(m_W)=173.8$~GeV and $\overline{m}_b(m_W)=3.06$~GeV have been used. *\,The $B\to X_s \ell^+\ell^-$ bounds on $\mathrm{Re}[\kappa_{RR,LRb}]$ are valid in the $\mathrm{Im}[\kappa_{RR,LRb}]=0$ limit; see text for details. Accompanying plot serves for visual comparison of the presently allowed intervals.}
\label{tab:bounds}
\end{table}

The first column shows bounds obtained from $B_{d,s}-\bar{B}_{d,s}$ mixing as analyzed in Ref.~\cite{Drobnak:2011wj}, while the last column corresponds to combined bounds from all three observables. For the later we also present the results when the matching scale is set to $\mu=m_W$ to check the scale dependence of our results. We can see that the bounds obtained change significantly only in the case of $\kappa_{LRb}$ where lowering the scale to $\mu=m_W$ loosens the bounds by almost an order of magnitude. We have also checked that the $B\to X_s \gamma$ bounds agree nicely with those obtained in Ref.~\cite{Grzadkowski:2008mf}.

\subsection{Imaginary parts}
We have shown in Ref.~\cite{Drobnak:2011wj} that imaginary parts of primed Wilson coefficients can affect the CP violating phase in $B_{d,s}-\bar B_{d,s}$ mixing and nonzero values were found to improve the global fit of~\cite{Lenz:2010gu}. 
To constrain imaginary parts of the remaining four operators, which do not contribute with new phases in $B_{d,s}-\bar B_{d,s}$ mixing, we consider the direct CP asymmetry in $B\to X_s \gamma$ for which the current world average experimental value reads \cite{Asner:2010qj}
\begin{eqnarray}
A_{X_s \gamma}=\frac{\Gamma(\bar{B}\to X_s\gamma)-\Gamma(B\to X_{\bar{s}}\gamma)}{\Gamma(\bar{B}\to X_s\gamma)+\Gamma(B\to X_{\bar{s}}\gamma)} = -0.012 \pm 0.028\,.
\end{eqnarray} 
Based on the recent analysis of this observable in Ref.~\cite{Benzke:2010tq} we obtain the following semi-numerical formula
\begin{eqnarray}
A_{X_s \gamma}&=&0.006+0.039(\tilde{\Lambda}_{17}^u-\tilde{\Lambda}_{17}^c) \nonumber\\
&+&\Big[0.008+0.051 (\tilde{\Lambda}_{17}^u-\tilde{\Lambda}_{17}^c)\Big]\mathrm{Re}[\delta C_7(2m_W)]
+\Big[0.012(\tilde{\Lambda}_{17}^u- \tilde{\Lambda}_{17}^c)+0.002\Big]\mathrm{Re}[\delta C_8(2m_W)]\nonumber\\
&+&\Big[-0.256+0.264 \tilde{\Lambda}_{78}-0.023 \tilde{\Lambda}_{17}^u-2.799 \tilde{\Lambda}_{17}^c\Big]\mathrm{Im}[\delta C_7(2m_W)]\nonumber\\
&+&\Big[-0.668 \tilde{\Lambda}_{17}^c-0.005 \tilde{\Lambda}_{17}^u-0.563 \tilde{\Lambda}_{78}+0.135\Big]\mathrm{Im}[\delta C_8(2m_W)]\,.
\end{eqnarray}
The estimated intervals for hadronic parameters $\tilde{\Lambda}_{17}^u$, $\tilde{\Lambda}_{17}^c$ and $\tilde{\Lambda}_{78}$  as specified in Ref.~\cite{Benzke:2010tq} dominate the theoretical uncertainty making it sufficient to use a LO QCD analysis in the perturbative regime. Thus, in addition to the numerical parameters specified in~\cite{Benzke:2010tq}, we have used the LO QCD running for $\delta C_{7,8}$ in this observable. 

Performing a combined analysis of all considered bounds on the real and imaginary parts of individual $\kappa_i$ in which we marginalize the hadronic parameters entering $A_{X_s\gamma}$ within the allowed intervals, we can obtain the allowed regions in the complex plain of $({\rm Re}[\kappa_i],{\rm Im}[\kappa_i])$.  
\begin{figure}[h]
\begin{center}
\includegraphics[scale=0.65]{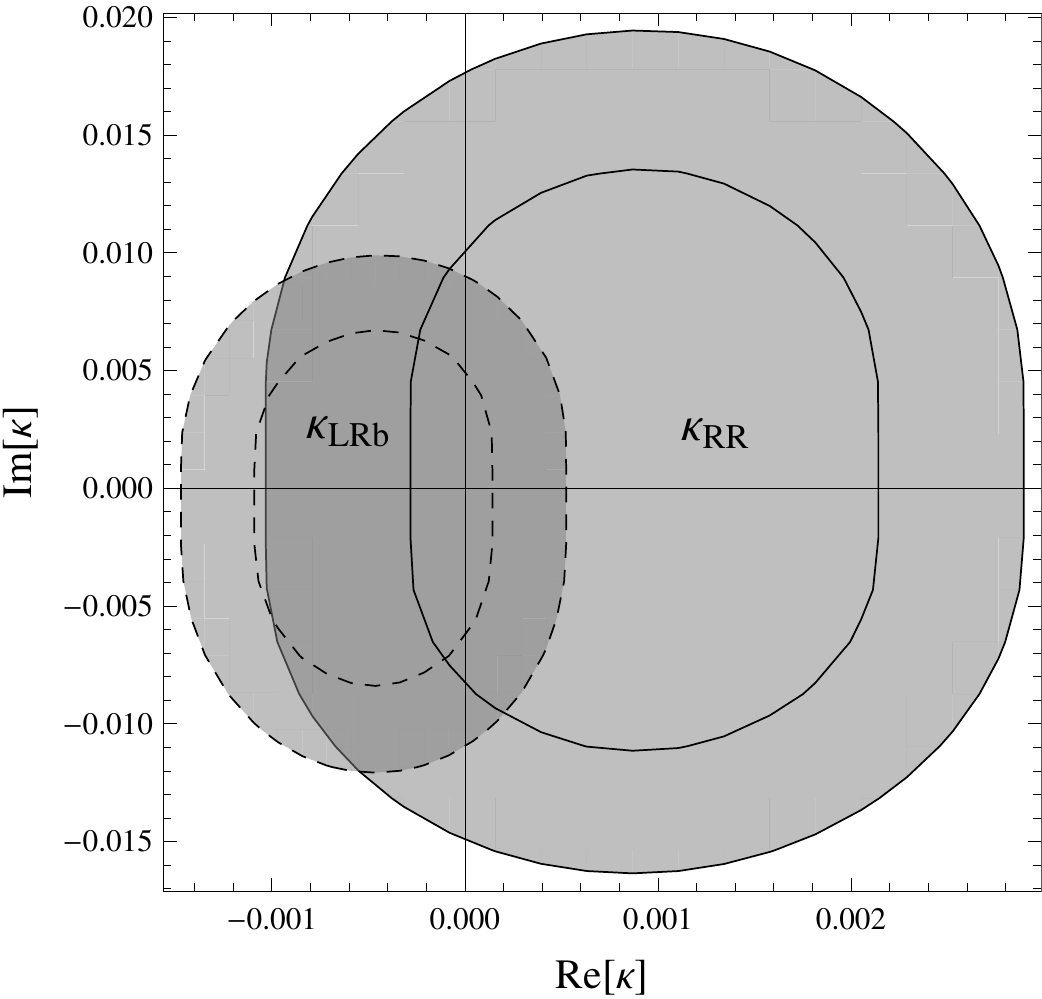}
\caption{$68\%$ and $95\%$ C.L. allowed regions in the $\kappa_{RR}(2m_W)$ (solid) and $\kappa_{LRb}(2m_W)$ (dashed) complex plain. The constraints are dominated by ${\rm Br}[{B}\to X_s \gamma]$ and $A_{X_s\gamma}$.}
\label{fig:CP}
\end{center}
\end{figure}

As already argued, the imaginary part of $\kappa_{LL}$ does not contribute to the $\delta C_i$ and thus remains unconstrained. It also turns out that due to the large hadronic uncertainties, the imaginary parts of $\kappa^{(\prime)}_{LRt}$, $\kappa^{\pr(\pr\pr)}_{LL}$ remain largely unconstrained by $A_{X_s \gamma}$ and $B_{d,s}-\bar B_{d,s}$ mixing observables still provide the strongest constraints~\cite{Drobnak:2011wj} (except for ${\rm Im}[\kappa_{LRt}]$ which again remains unconstrained). On the contrary, constraints on the imaginary parts of $\kappa_{RR}$ and $\kappa_{LRb}$ reach per-cent level, as can be seen in Fig.~\ref{fig:CP}.
Finally we note that in absence of the long-distance effects on NP contributions considered in~\cite{Benzke:2010tq}, $A_{X_s \gamma}$ would exhibit an even greater sensitivity to the imaginary parts of $\delta C_{7,8}$~\cite{Barbieri:2011fc}, thus we consider our derived bounds as conservative.

\subsection{Comparison with direct constraints}
\begin{figure}[h]
\begin{center}
\includegraphics[scale= 0.47]{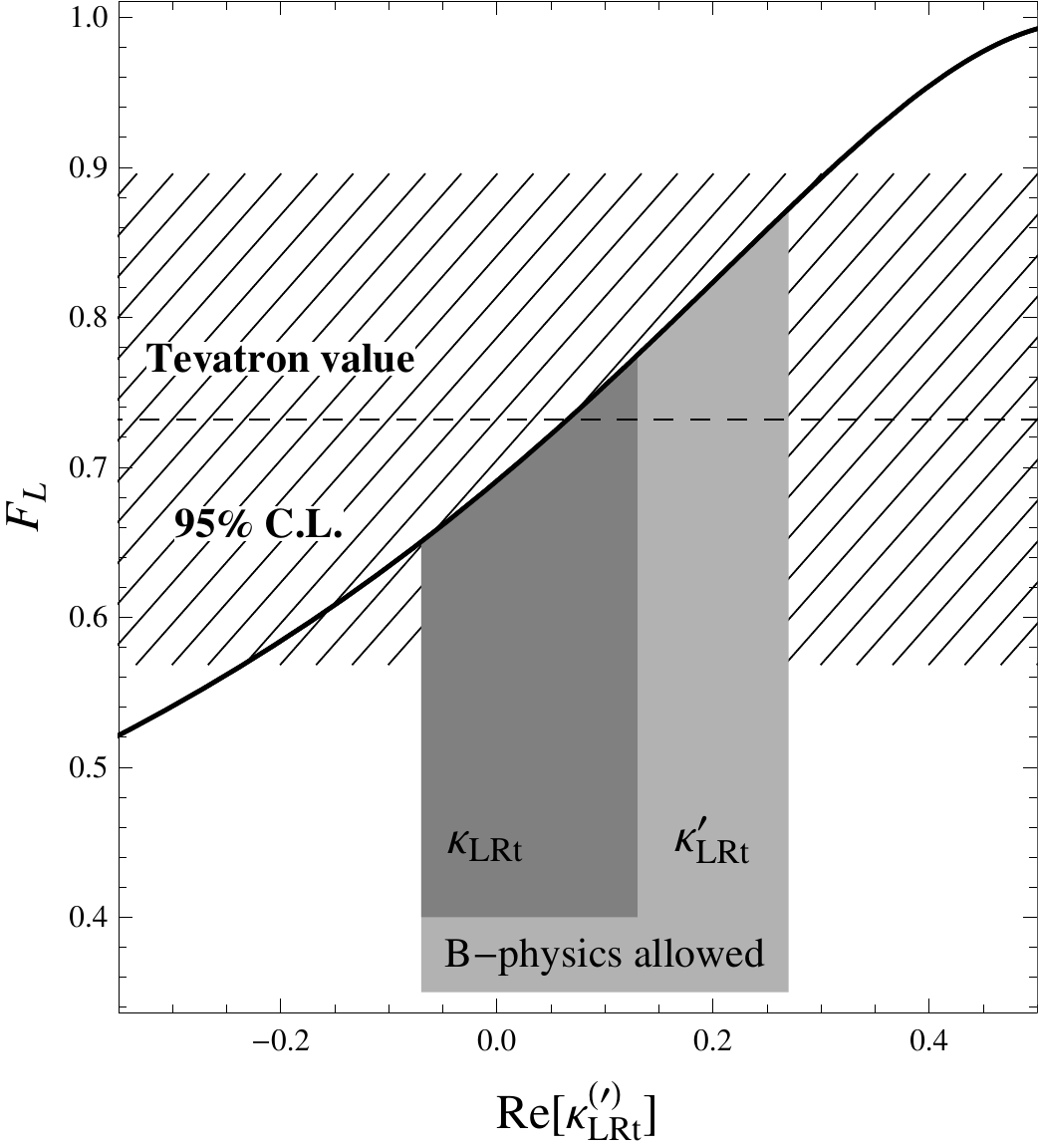}\hspace{0.5cm}
\includegraphics[scale= 0.54]{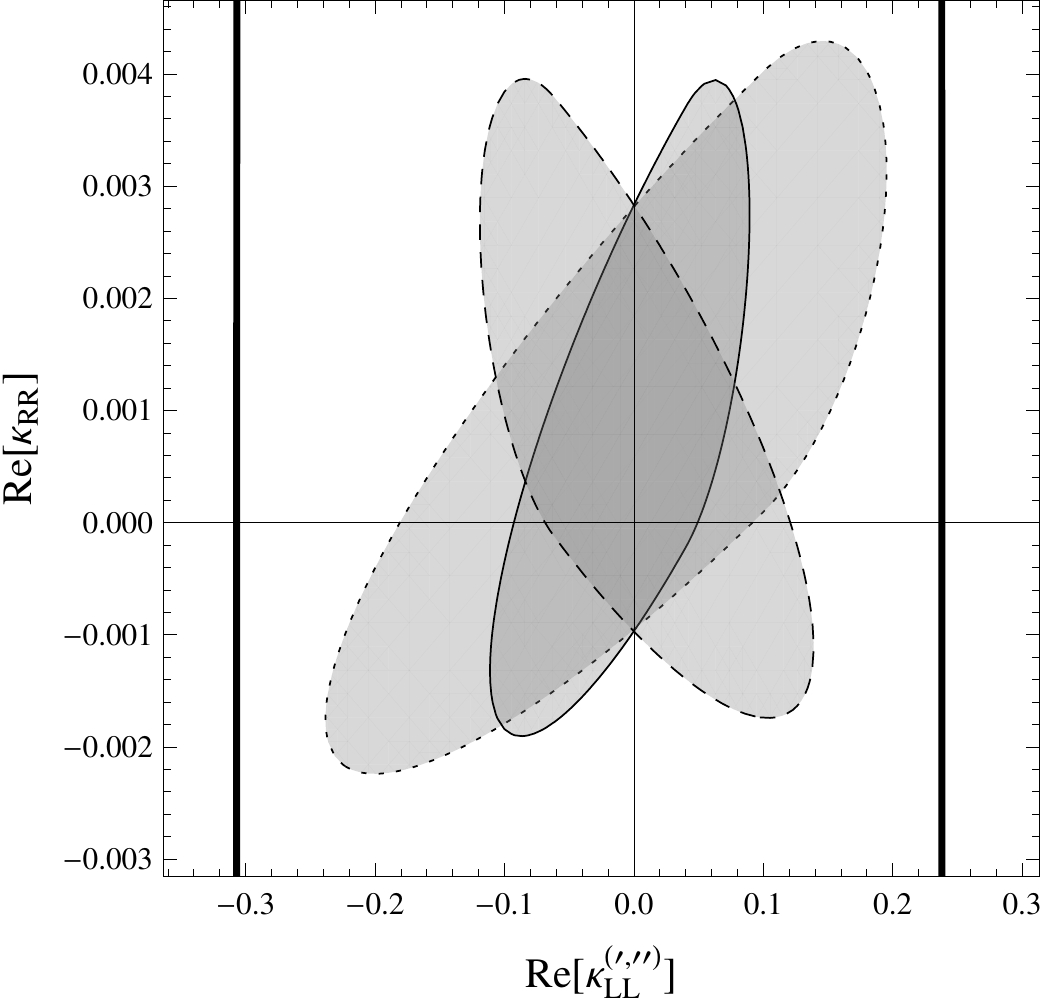}\hspace{0.5cm}
\includegraphics[scale=0.54]{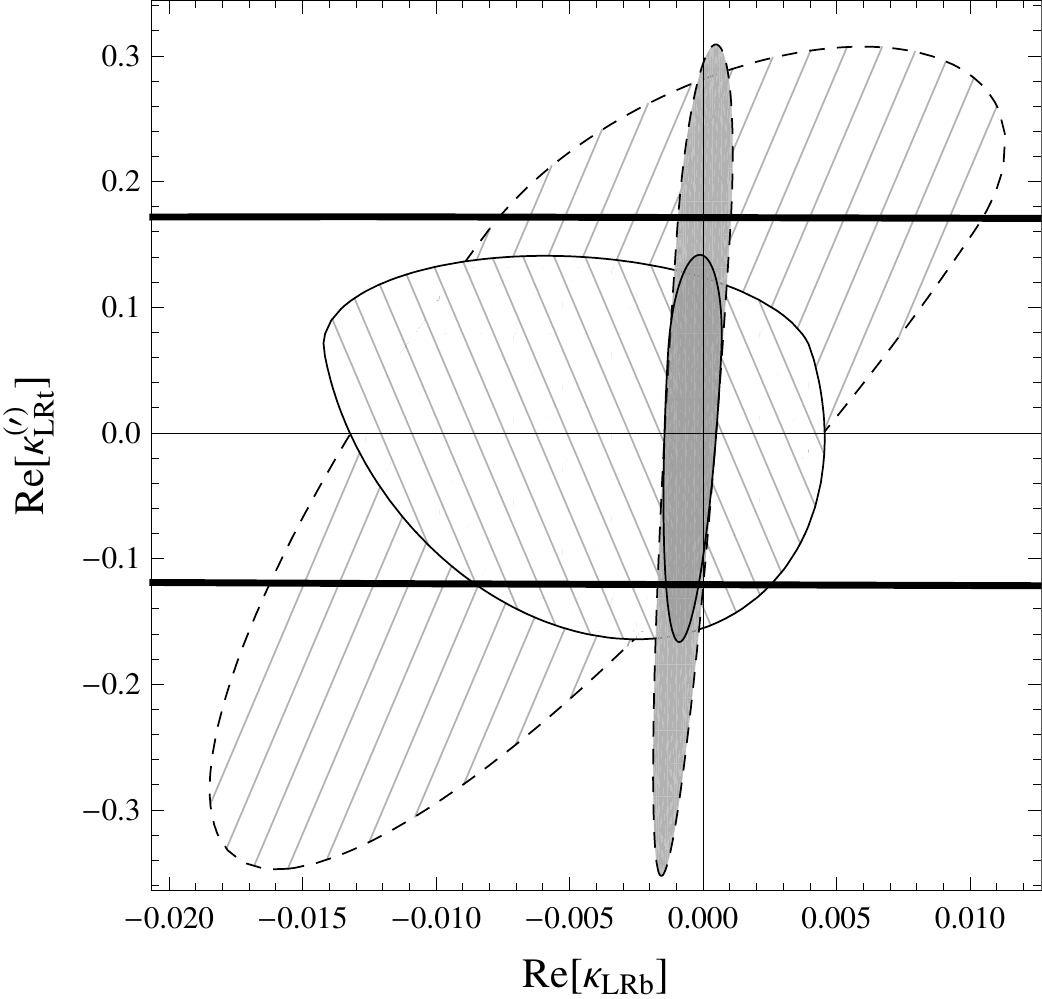}
\caption{{\bf Left}: Longitudinal helicity fraction as a function of the real part of $\kappa_{LRt}^{(\pr)}$ obtained from \cite{Drobnak:2010ej}. Also shown is the latest combined Tevatron measurement~\cite{newHelicity} and allowed intervals for $\kappa_{LRt}^{(\pr)}(2m_W)$ as given in Tab.~\ref{tab:bounds}. {\bf Middle}: $95\%$ C.L. allowed regions in the $\kappa_{RR}$ - $\kappa_{LL}$ (solid), $\kappa_{LL}^{\prime}$ (dashed), $\kappa_{LL}^{\pr\pr}$ (dotted) plane. Matching scale is set to $\mu=2 m_W$. {\bf Right}: $95\%$ C.L. allowed regions in the $\kappa_{LRb}$ - $\kappa_{LRt}$ (solid), $\kappa_{LRt}^{\prime}$ (dashed) plane. Matching scale is set to $\mu=2 m_W$ (full regions) and $\mu=m_W$ (hatched regions). Thick black lines correspond to bands of $95\%$ C.L. allowed regions given by direct constraints from Ref.~\cite{AguilarSaavedra:2011ct}.}
\label{fig:2d1}
\end{center}
\end{figure}

Anomalous $tWb$ couplings can be studied directly at colliders, namely in the main top quark decay channel ($t\to b W$) as well as in the single top production since in the SM both processes proceed through weak interactions. Contrary to indirect studies of $tWb$ couplings in $B$ physics, the most general parametrization of the $tWb$ vertex of the form
\begin{eqnarray}
{\cal L}_{tWb} ={\cal L}_{tWb}^{\mathrm{SM}} -\frac{g}{\sqrt{2}}\bar{b}\Big[(V_L P_L + V_R P_R)\gamma^{\mu}+\frac{\ii \sigma^{\mu\nu}q_{\nu}}{m_W}(G_L P_L + G_R P_R)\Big]tW_{\mu}\,,
\end{eqnarray}
is sufficient to a very good approximation, where $P_{L,R}=1/2(1\mp\gamma^5)$. It is straightforward to identify the anomalous couplings $V_{R,L}$ and $G_{R,L}$ as generated by our seven operators in Eq. (\ref{eq:ops1})
\begin{equation}
v_{L}= V_{tb}^*\kappa_{LL}^{(\pr,\pr\pr)*}\,, \hspace{0.3cm} v_R =V_{tb}^* \kappa_{RR}^*\,, \hspace{0.3cm} g_L=V_{tb}^* \kappa_{LRb}^*\,,\hspace{0.3cm} g_R=V_{tb}^*\kappa_{LRt}^{(\prime)}\,,
\end{equation}
enabling us to compare recent direct bounds obtained from the Tevatron data~\cite{Drobnak:2010ej, AguilarSaavedra:2011ct} with our indirect constraints. The comparison is shown in Fig.~\ref{fig:2d1}.

The first plot shows the longitudinal helicity fraction $F_L$ as a function of the real parts of the anomalous couplings $\kappa_{LRt}^{(\pr)}$ obtained from \cite{Drobnak:2010ej} and the latest combined measurements from Tevatron with the $95\%$ C.L. band \cite{newHelicity}. We find the combined indirect constraints to be more stringent than direct bounds coming from helicity fraction measurements. For other operators this conclusion is even more pronounced.

Having more than one observable at disposal we can also consider pairs of operators contributing simultaneously and obtain $95\%$ C.L. allowed regions in the corresponding planes as shown in the second and third plot of Fig.~\ref{fig:2d1}. The full and hatched shapes are obtained from indirect bounds while thick black line marks the border of the region obtained from helicity fraction and single top production analysis of Ref.~\cite{AguilarSaavedra:2011ct}. In particular both vertical lines in the second graph are due to  single top production measurements, as helicity fractions are insensitive to a change in the left-handed couplings. Also, the bottom thick line in the third graph is due to constraints from single top production, since as indicated in the first graph, the negative values of $\kappa_{LRt}^{(\pr)}$ are better constrained indirectly than from helicity fractions. It is apparent that indirect constraints on the real parts of $\kappa_{RR}$ and $\kappa_{LRb}$ are at present much more stringent than direct constraints.


\section{Predictions}
Having derived bounds on anomalous $\kappa_j$ couplings, it is interesting to study to what extent these can still affect other rare $B$ decay observables.  Analyzing one operator at a time we set the matching scale to $\mu=2m_W$.

Turning again to the semi-numerical formulae given in Ref.~\cite{DescotesGenon:2011yn}, we first consider  the branching ratio $\mathrm{Br}[\bar B_s\to\mu^+\mu^-]$ for which CDF's latest analysis yields~\cite{Aaltonen:2011fi}
\begin{eqnarray}
4.6\times 10^{-9}<\mathrm{Br}[\bar B_s\to\mu^+\mu^-]<3.9\times 10^{-8}\,,\hspace{0.5cm} \text{at $90\%$ C.L.}\,.
\end{eqnarray}
and the differential forward-backward asymmetry $A_{\mathrm{FB}}(q^2)$ in the $\bar{B}_d\to \bar{K}^*\ell^+\ell^-$ decay, for which the latest measurement of LHCb has recently been presented in Ref.~\cite{newAsymm}.
Finally, following Ref.~\cite{Altmannshofer:2009ma} we analyze the allowed effects of $\kappa_i$ on the branching ratios $\mathrm{Br}(B\to K^{(*)}\nu\bar{\nu})$, which are expected to become experimentally accessible at the super-B factories~\cite{SuperB}.

\begin{figure}[h]
\begin{center}
\includegraphics[scale= 0.513]{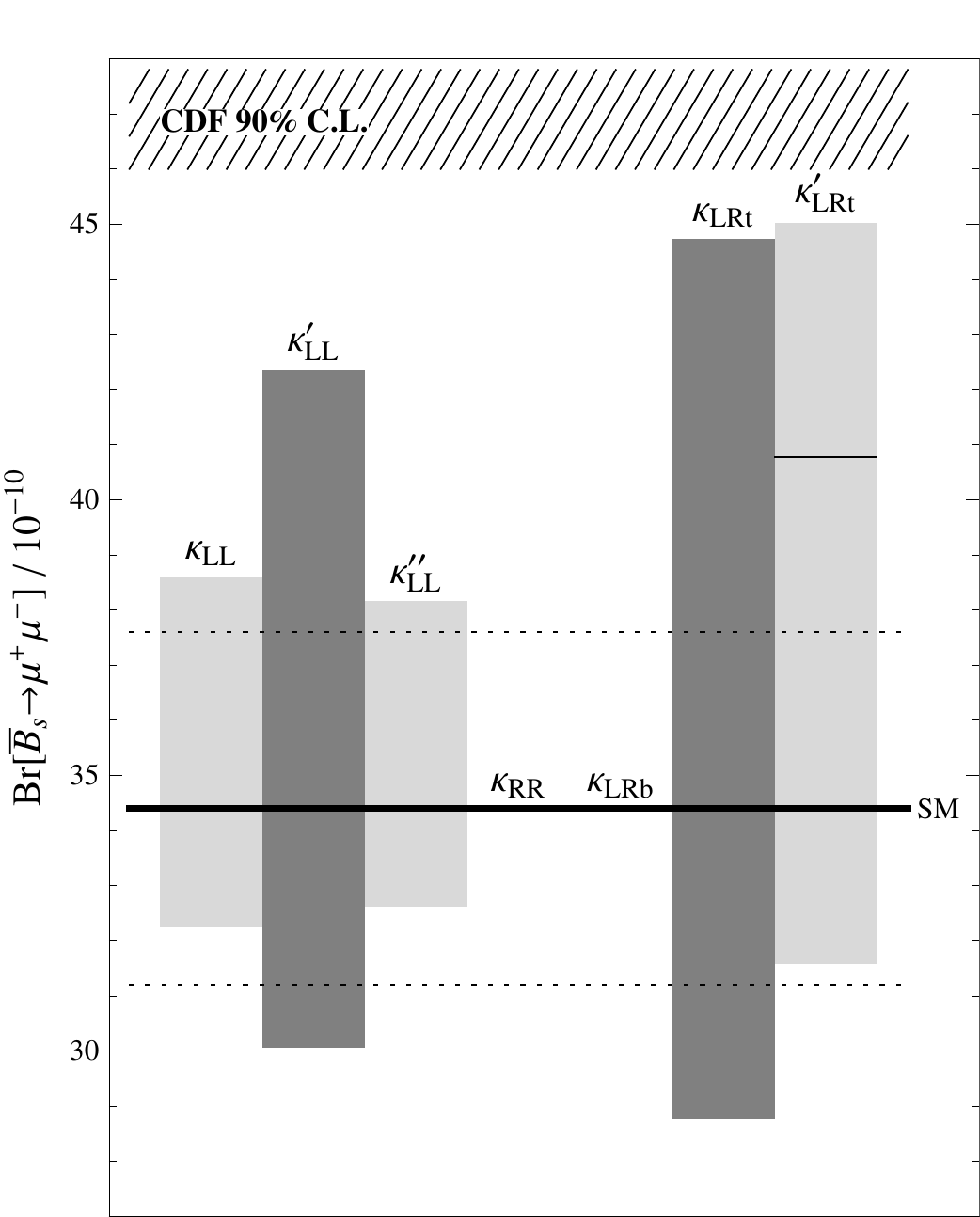}\hspace{0.5cm}
\includegraphics[scale=0.48]{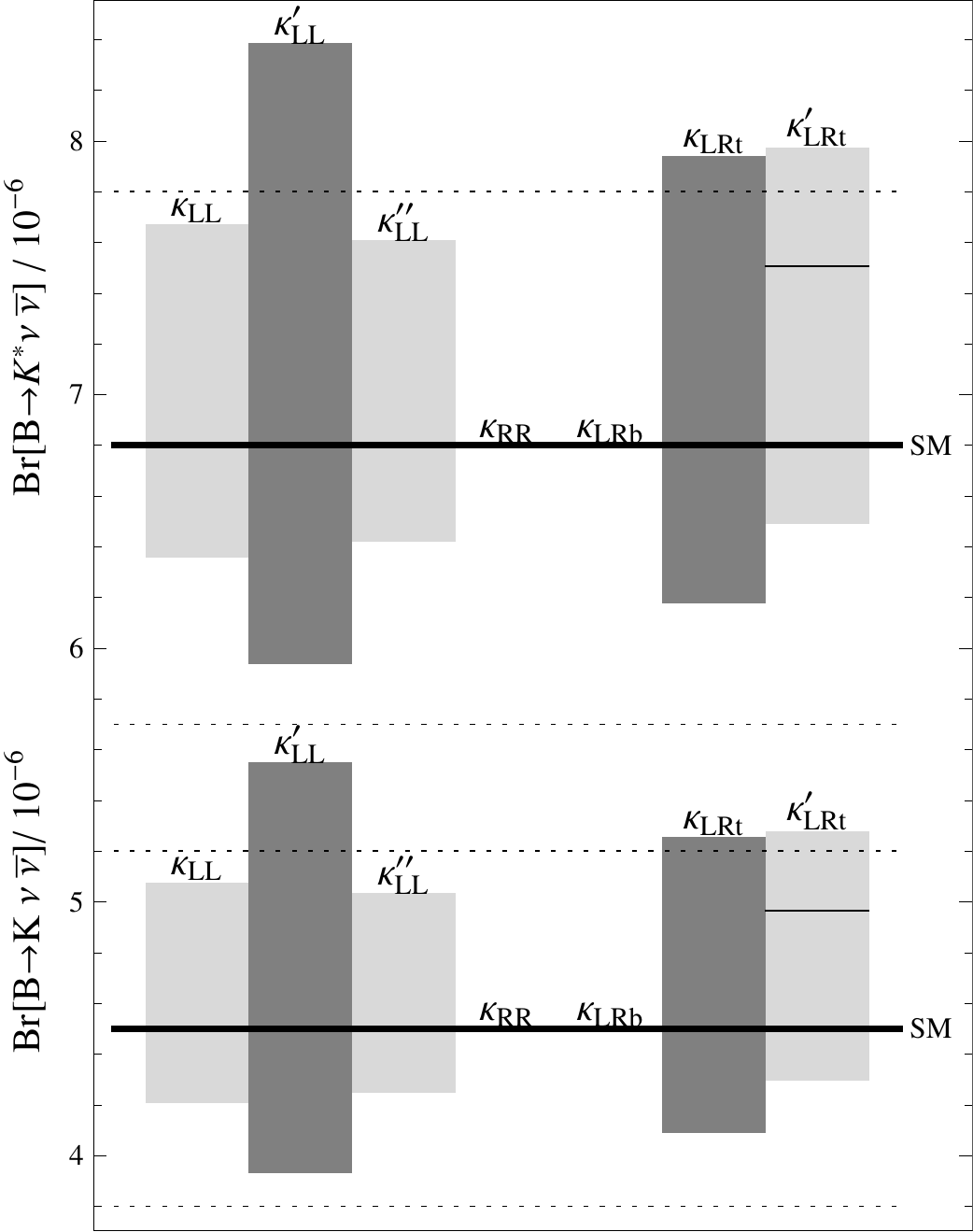}\hspace{0.5cm}
\caption{Ranges of values for branching ratios obtained as anomalous couplings are varied within the $95\%$ C.L. intervals given in Tab.~\ref{tab:bounds}. Line in the $\kappa_{LRt}^{\prime}$ bar marks the value obtained when the coupling is set to the central fitted value given by the $B_{d,s}-\bar B_{d,s}$ mixing analysis. We also show the SM predictions (black) with $1\sigma$ theoretical uncertainty band (dotted) and for the muonic decay channel the lower end of the experimental $90\%$ C.L. interval from \cite{Aaltonen:2011fi}.}
\label{fig:predict1}
\end{center}
\end{figure}

\begin{figure}[h]
\begin{center}
\includegraphics[scale= 0.51]{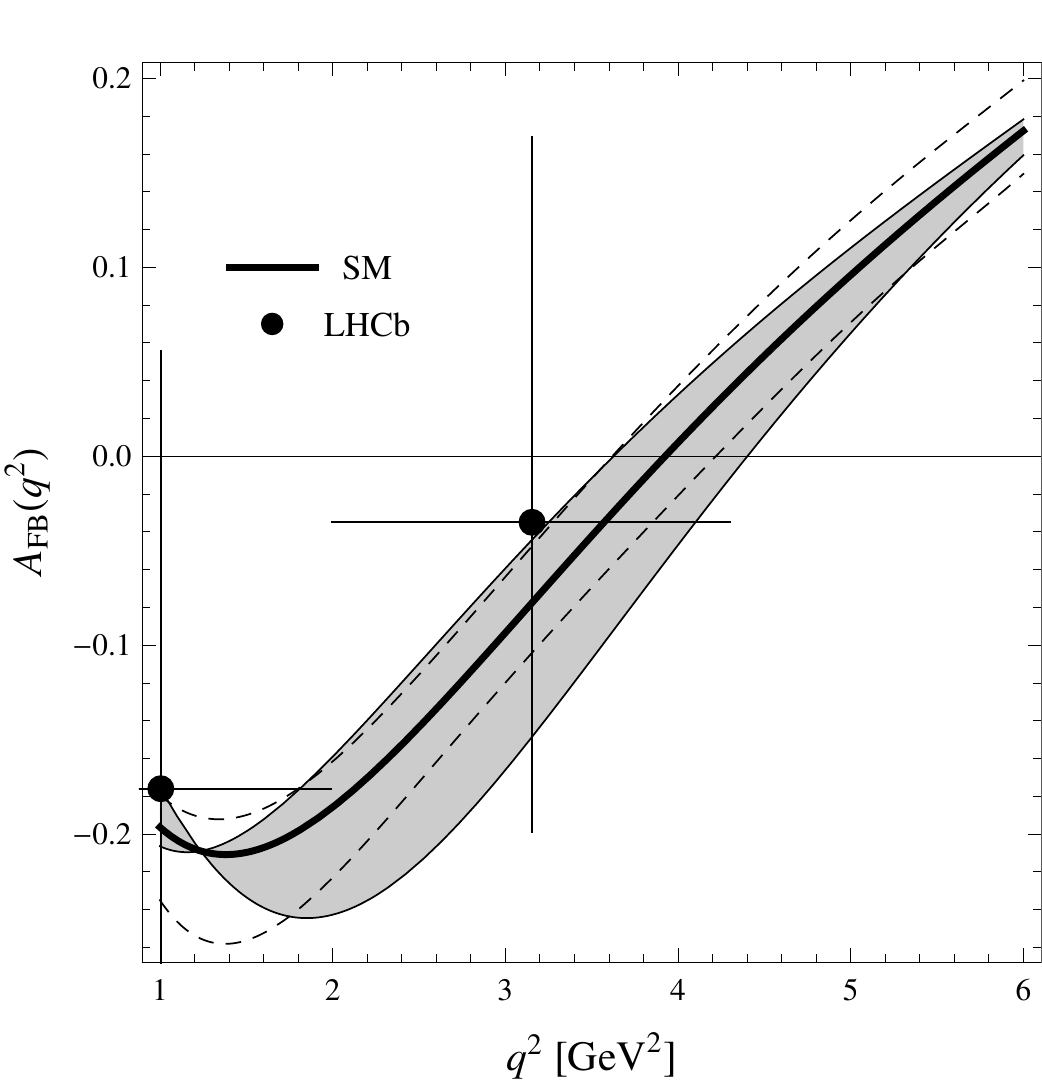}\hspace{0.5cm}
\includegraphics[scale= 0.51]{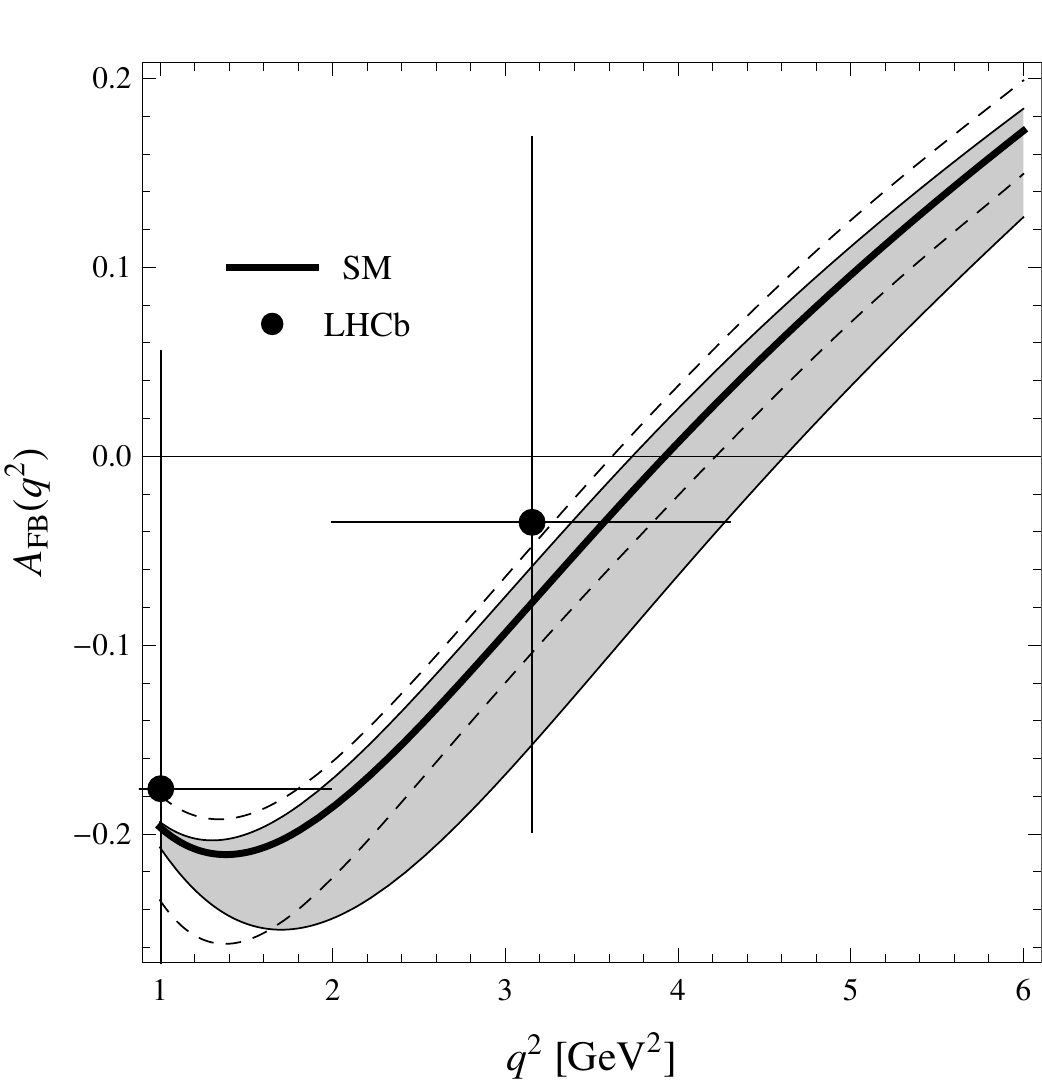}
\caption{$A_{\mathrm{FB}}(q^2)$ band obtained when varying real parts of $\kappa_{LL}^{\pr\pr}$ (left) and $\kappa_{RR}$ (right) within the 95\% C.L. interval given in Tab.~\ref{tab:bounds}. Also presented are the SM predicted central value (black) with $1\sigma$ theoretical uncertainty band (dotted) and the latest measured points with experimental errors given in Ref.~\cite{newAsymm}.  }
\label{fig:predict2}
\end{center}
\end{figure}

We present our findings in Figs.~\ref{fig:predict1} and~\ref{fig:predict2}. The effects of anomalous couplings $\kappa_j$ on all branching ratios are similar. There is a slight tenancy of anomalous $\kappa_j$ couplings to increase the predictions compared to the SM values at the level of the present theoretical uncertainties, with the exception of $\kappa_{RR}$ and $\kappa_{LRb}$ of which effects are negligible. In particular, none of the contributions can accommodate the recent CDF measurement of ${\rm Br}[\bar B_s \to \mu^+\mu^-]$ at the $95\%$ C.L.\,, while a possible future measurement at the level of the SM could significantly constrain the $\kappa'_{LL}$ and $\kappa^{(\prime)}_{LRt}$ contributions. Furthermore we find that the forward-backward asymmetry $A_{\mathrm{FB}}(q^2)$ can still be somewhat effected by $\kappa_{LL}^{\pr\pr}$ and $\kappa_{RR}$, for which we present the bands obtained when varied within the $95\%$ C.L. intervals in Fig.~\ref{fig:predict2}. While not sensitive at the moment, in the near future, improved measurements by the LHCb experiment could possibly probe such effects. On the other hand, the contributions of other anomalous couplings all fall within the theoretical uncertainty bands around the SM predicted curve.

\section{Conclusions}
We have investigated contributions of  anomalous $tWb$ couplings in flavor changing neutral current mediated $\Delta B=1$ processes within an effective field theory framework assuming minimal flavor violation. Having computed contributions to the inclusive $B \to X_s \ell^+ \ell^-$ decay rate to order $1/\Lambda^2$, and combining them with the modifications of the  $ B\to X_s \gamma$ and also $B_{d,s}- \bar B_{d,s}$ observables, we have determined the indirect bounds on the real and imaginary parts of the anomalous $tWb$ couplings. For most of the considered effective operators, these indirect bounds are at present much stronger than the direct constraints coming from the $t\to b W$ helicity fraction measurements, angular asymmetries and single top production at the Tevatron and the LHC.
In particular, we are able for the first time to constrain the imaginary parts of most of the anomalous couplings already at order $1/\Lambda^2$. Taking into account these bounds, we have predicted the still allowed effects of the anomalous $tWb$ interactions on the branching ratio of the $B_s \to \mu^+ \mu^-$ decay, the forward-backward asymmetry in $B \to K^* \ell^+ \ell^-$, as well as the branching ratios of $B \to K^{(*)} \nu \bar \nu$ decays. The better knowledge of these and other recently proposed~\cite{Bobeth:2007dw} observables in the future could further constrain some of the anomalous couplings. 

\begin{acknowledgments}
We thank Bohdan Grzadkowski and Mikolaj Misiak for useful correspondence 
and for positive verification of our corrected result for the function
$f_{7}^{g_L}$ in eq. (12) of Ref.~\cite{Grzadkowski:2008mf}.
J.F.K. acknowledges insightful conversations with Jure Zupan, Joaquim Matias, S\'ebastien Descotes-Genon, and Gino Isidori. This work is supported in part by the Slovenian Research Agency, by the National Science Foundation under Grant No. 1066293 and the hospitality of the Aspen Center for Physics.
\end{acknowledgments}

\newpage
\appendix
\section{Feynman Rules}\label{sec:app1}
The relevant Feynman rules with anomalous couplings are shown in Tab.~\ref{tab:feyns}.
We have used the following abbreviations 
\begin{eqnarray}
v_R &=& \kappa_{RR} \delta_{3i}\delta_{3j}\,,\\
\tilde{v}_R&=&\frac{c_W}{s_W}v_R\,,\\ 
v_L &=& \kappa_{LL}\delta_{3i}+\kappa_{LL}^{\pr}\delta_{3j}+\kappa_{LL}^{\pr\pr}\delta_{3i}\delta_{3j}\,,\\
\tilde{v}_L&=&\frac{c_W^2-s_W^2}{2 c_W s_W} v_L-\frac{1}{2c_Ws_W}\Big(\kappa_{LL}^*\delta_{3i}+\kappa_{LL}^{\pr*}\delta_{3j}+\kappa_{LL}^{\pr\pr*}\frac{V_{ib}V_{tj}}{V_{ij}}\Big)\,,\\
g_R &=& -\kappa_{LRb}\,,\\
g_L &=& -\kappa_{LRt}^* \delta_{3i}-\kappa_{LRt}^{\pr*}\delta_{3i}\delta_{3j}\,,
\end{eqnarray}
where $i,j$ denote flavor indices.

\begin{table}[h!]
\begin{tabular}{m{2.0cm}l|m{2.0cm}l}
\includegraphics[scale= 0.5]{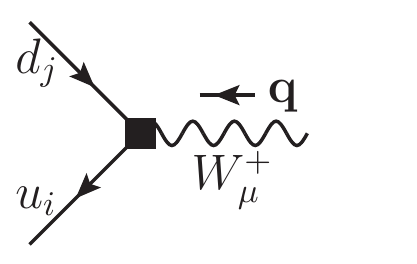}
 &$-\frac{\ii g}{\sqrt{2}}V_{ij}\Big[\gamma^{\mu}v_{R,L} +\frac{\ii \sigma^{\mu\nu}q_{\nu}}{m_W}g_{R,L}\Big]P_{R,L}$ &
\includegraphics[scale= 0.5]{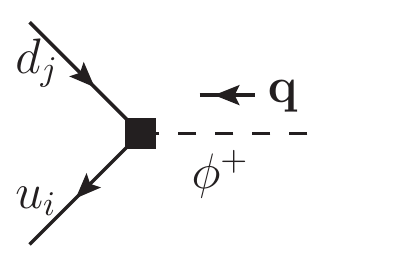}
 &$-\frac{\ii g}{\sqrt{2}}V_{ij}\frac{\gs{q}}{m_W}(-v_{R,L}) P_{R,L} $\\
 \includegraphics[scale= 0.5]{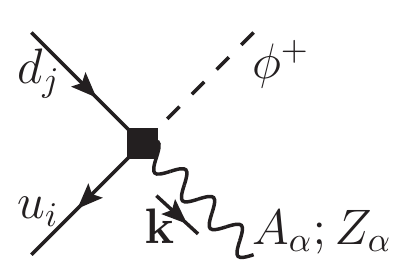}
 &$\begin{array}{l}-\frac{\ii g}{\sqrt{2}} V_{ij}\frac{e}{m_W}\Big[\{v_{R,L};\tilde{v}_{R,L} \}\gamma^{\mu}\\
 +\{1;\frac{c_W}{s_W}\}(-g_{R,L})\frac{\ii \sigma^{\alpha\mu}k_{\mu}}{2 m_W}\Big]P_{R,L}\end{array}$&
\includegraphics[scale= 0.5]{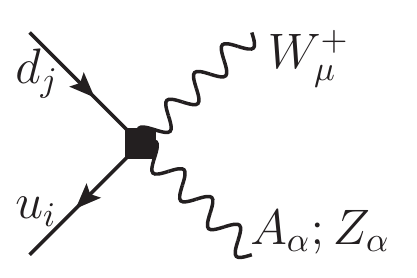}
 &$-\frac{\ii g}{\sqrt{2}}V_{ij} e\Big[\{1;\frac{c_W}{s_W}\}(-g_{R,L})\frac{\ii \sigma^{\mu\alpha}}{m_W}\Big]P_{R,L}$\\\hline
 \includegraphics[scale= 0.5]{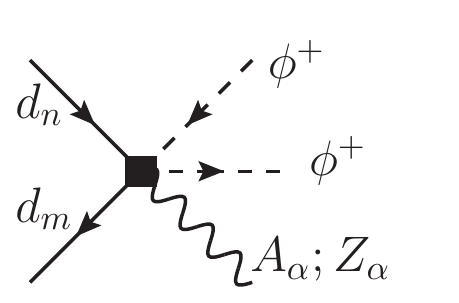}
 &$\begin{array}{l}\ii\big(\frac{g}{\sqrt{2}}\big)^2 \frac{e}{m_W^2}V_{tm}^*V_{tn}\{1;\frac{c_W^2-s_W^2}{2c_W s_W}\}\gamma^{\alpha}P_L\\
 \times(\kappa_{LL}+\delta_{3n}\kappa_{LL}^{\pr\pr})\end{array}$&
\includegraphics[scale= 0.5]{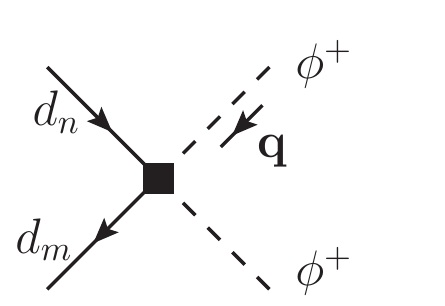}
 &$\begin{array}{l}-\ii\big(\frac{g}{\sqrt{2}}\big)^2V_{tm}^* V_{tn}\frac{ \gs{q}}{m_W^2}{P}_L\\\times(\kappa_{LL}+\delta_{3n}\kappa_{LL}^{\prime\prime})\end{array}$\\
\includegraphics[scale= 0.5]{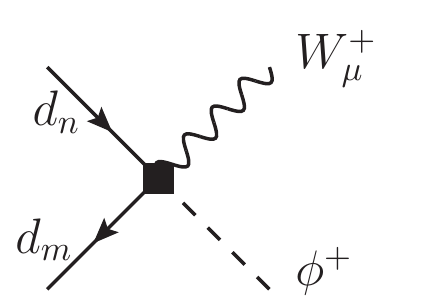}
 &$\begin{array}{l}\ii \big(\frac{g}{\sqrt{2}}\big)^2V_{tm}^* V_{tn}\frac{1}{m_W}\gamma^{\mu}{P}_L\\ \times(\kappa_{LL}+\delta_{3n}\kappa_{LL}^{\prime\prime})\end{array}$&
 \includegraphics[scale= 0.5]{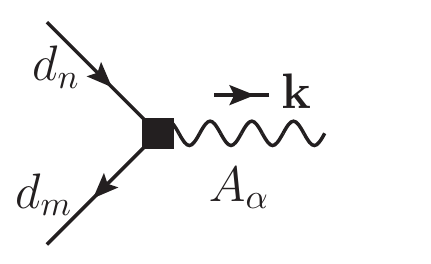}
 &$\ii e\,\kappa_{LRb}\delta_{3m}\delta_{3n}\frac{\ii \sigma^{\alpha\mu}k_{\mu}}{2m_W}P_R$\\\hline
\includegraphics[scale= 0.5]{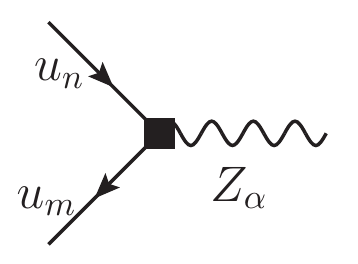}
 &$\begin{array}{l}-\ii e \frac{1}{2s_Wc_W}\gamma^{\alpha}P_L\\
 \times\big(\delta_{3m}\delta_{3n}\kappa_{LL}+V_{mb}V_{nb}^*(\kappa_{LL}^{\pr}+\delta_{3m}\kappa_{LL}^{\pr\pr})\big)\end{array}$&
\includegraphics[scale= 0.5]{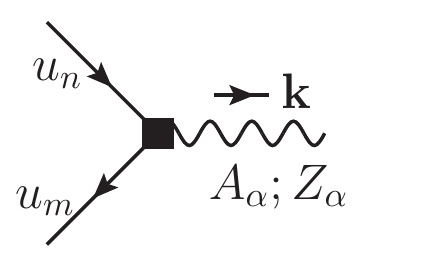}
 &$\begin{array}{l}-\ii e\{1;\frac{c_W}{s_W}\} \frac{\ii \sigma^{\alpha\nu}k_{\nu}}{2 m_W}P_R \\ 
 \times\big(\delta_{3m}\delta_{3n} \kappa_{LRt}+V_{mb}\delta_{3n}\kappa_{LRt}^{\pr}\big)\end{array}$ \\
\end{tabular}
\caption{ Feynman rules for the relevant anomalous vertices. ${P}_{L,R}=(1\mp\gamma^5)/2$, indicies $i,j$ and $m,n$ label quark flavor.}
\label{tab:feyns}
\end{table}
\vfill

\section{Analytical formulae}\label{sec:app2}
In this Appendix we present analytical expressions for functions $f_i^{(j)}$ and $\tilde{f}_i^{(j)}$ defined in Eq. (\ref{eq:fs}). For shorter notation we further decompose
\begin{eqnarray*}
f_{9}^{(j)}=g^{(j)}-\frac{1}{4s_W^2} h^{(j)}\,,\hspace{0.5cm}f_{10}^{(j)}=\frac{1}{4 s_W^2}h^{(j)}\,,\hspace{0.5cm} f^{(j)}_{\nu\bar{\nu}}=\frac{1}{4s_W^2}k^{(j)}\,.
\end{eqnarray*}
Below we give all nonzero contributions
\begin{eqnarray}
f_7^{(LL)} &=&\tilde{f}_7^{(LL)}=f_7^{(LL\pr\pr)}= \frac{22 x^3-153x^2+159x-46}{72 (x-1)^3}+\frac{3x^3-2x^2}{4(x-1)^4}\log x\\
f_7^{(LL\pr)}&=&-\frac{8x^3+5x^2-7x}{24(x-1)^3}+\frac{3x^3-2x^2}{4(x-1)^4}\log x \\
f_7^{(RR)}&=&\frac{m_t}{m_b}\Big[\frac{-5x^2+31x-20}{12 (x-1)^2}+\frac{2x-3x^2}{2(x-1)^3}\log x\Big]\label{eq:rr}\\
f_7^{(LRb)}&=&\frac{m_W}{m_b}\Big[-\frac{x}{2}\log \frac{m_W^2}{\mu^2}+\frac{6x^3-31x^2+19x}{12(x-1)^2}+\frac{-3x^4+16x^3-12x^2+2x}{6(x-1)^3}\log x\Big] \label{eq:lrb}\\
f_7^{(LRt)}&=&\frac{m_t}{m_W}\Big[\frac{1}{8}\log\frac{m_W^2}{\mu^2}+\frac{-9x^3+63x^2-61x+19}{48(x-1)^3}+\frac{3x^4-12x^3-9x^2+20x-8}{24(x-1)^4}\log x\Big]\\
\tilde{f}_7^{(LRt)}&=&\tilde{f}_7^{(LRt\prime)}=\frac{m_t}{m_W}\Big[\frac{-3x^3+17x^2-4x-4}{24(x-1)^3}+\frac{2x-3x^2}{4(x-1)^4}\log x\Big]\\
f_7^{(LRt\prime)}&=&\frac{m_t}{m_W}|V_{tb}|^2\Big[\frac{-x^2-x}{8(x-1)^2}+\frac{x^2\log x}{4(x-1)^3}\Big]\\
f_8^{(LL)}&=&\tilde{f}_8^{(LL)}=f_8^{(LL\pr\pr)}=\frac{5 x^3-9 x^2+30 x-8}{24 (x-1)^3}-\frac{3 x^2 \log x}{4 (x-1)^4}\\
f_8^{(LL\pr)}&=&\frac{-x^3+5x^2+2x}{8 (x-1)^3}-\frac{3 x^2 \log x}{4 (x-1)^4}\\
f_8^{(RR)}&=&\frac{m_t}{m_b}\Big[\frac{-x^2-x-4}{4 (x-1)^2}+\frac{3 x \log x}{2 (x-1)^3}\Big]\\
f_8^{(LRb)}&=&\frac{m_W}{m_b}\Big[\frac{x^2+5 x}{4 (x-1)^2}+\frac{2 x^3-6 x^2+x}{2 (x-1)^3}\log x\Big]\\
f_8^{(LRt)}&=&\frac{m_t}{m_W}\Big[\frac{3 x^2-13 x+4}{8 (x-1)^3}+\frac{5 x-2 }{4 (x-1)^4}\log x\Big]\\
\tilde{f}_8^{(LRt)}&=&\tilde{f}_8^{(LRt\pr)}=\frac{m_t}{m_W}\Big[\frac{x^2-5 x-2}{8 (x-1)^3}+\frac{3 x \log (x)}{4 (x-1)^4}\Big]\\
g^{(LL)}&=&\tilde{g}^{(LL)}=(-x-\frac{4}{3})\log\frac{m_W^2}{\mu^2}+\frac{250x^3-384x^2+39x+77}{108 (x-1)^3}\\
&+&\frac{-18x^5+48x^4-102x^3+135x^2-68x+8}{18(x-1)^4}\log x\\
g^{(LL\pr)}&=&(\frac{4}{9}-\frac{x}{2})\log\frac{m_W^2}{\mu^2}+\frac{125x^3-253x^2+138x -16}{36(x-1)^3}\\
&+&\frac{-9x^5+12x^4-48x^3+99x^2-59x+8}{18(x-1)^4}\log x - |V_{tb}|^2\frac{x}{2}\nonumber\\
\tilde{g}^{(LL\pr)}&=&\tilde{h}^{(LL\pr)}=\tilde{f}^{(LL\pr)}_{\nu\bar{\nu}}=-\frac{x}{2}\log\frac{m_W^2}{\mu^2}+\frac{x}{2}(1-\log x -|V_{tb}|^2)\\
g^{(LL\pr\pr)}&=&-\big(\frac{4}{3}+\frac{x}{2}+|V_{tb}|^2\frac{x}{2} \big)\log\frac{m_W^2}{\mu^2}
+ \frac{250x^3-384x^2+39x + 77}{108(x-1)^3}\\
&+&\frac{-9x^5+12x^4-48x^3+99x^2-59x+8}{18(x-1)^4}\log x - |V_{tb}|^2\frac{x}{2}\log x \nonumber \\
\end{eqnarray}%

\begin{eqnarray}
\tilde{g}^{(LL\pr\pr)}&=&\tilde{h}^{(LL\pr\pr)}=\tilde{f}^{(LL\pr\pr)}_{\nu\bar{\nu}}=|V_{tb}|^2\Big(-\frac{x}{2}\log\frac{m_W^2}{\mu^2}-\frac{x}{2}\log x \Big)\\
g^{(LRt)}&=&\tilde{g}^{(LRt)}=\frac{m_t}{m_W}\Big[\frac{-99x^3+136x^2+25x-50}{72(x-1)^3}+\frac{24x^3-45x^2+17x+2}{12(x-1)^4}\log x\Big]\\
g^{(LRt\pr)}&=&\frac{m_t}{m_W}|V_{tb}|^2\Big[\frac{x^2+3x-2}{8(x-1)^2}+\frac{x-2x^2}{4(x-1)^3}\log x\Big]\\
\tilde{g}^{(LRt\pr)}&=&\frac{m_t}{m_W}\bigg[\frac{-54x^3+59x^2+35x-34}{36(x-1)^3}+\frac{15x^3-27x^2+10x+1}{6(x-1)^4}\log x\\ 
&+& |V_{tb}|^2\Big[\frac{x^2+3x-2}{8(x-1)^2}+\frac{x-2x^2}{4(x-1)^3}\log x\Big]\bigg]\\
h^{(LL)}&=&\tilde{h}^{(LL)}=-(x+\frac{3}{2})\log\frac{m_W^2}{\mu^2}+\frac{11x-5}{4(x-1)}+\frac{-2x^3+x^2-2x}{2(x-1)^2}\log x\\
h^{(LL\pr)}&=&-\frac{x}{2}\log\frac{m_W^2}{\mu^2}+\frac{3x}{2(x-1)}-\frac{x^3+x^2+x}{2(x-1)^2}\log x-|V_{tb}|^2\frac{x}{2} \\
h^{(LL\pr\pr)}&=&-\big(\frac{3}{2}+\frac{x}{2}+|V_{tb}|^2\frac{x}{2}\big)\log\frac{m_W^2}{\mu^2}+\frac{11x-5}{4(x-1)}-\frac{x^3+x^2+x}{2(x-1)^2}\log x-|V_{tb}|^2 \frac{x}{2}\log x\\
h^{(LRt)}&=&\tilde{h}^{(LRt)}=\tilde{h}^{(LRt\pr)}=\frac{m_t}{m_W}\Big[-\frac{3x}{2(x-1)}+\frac{3x\log x}{2(x-1)^2}\Big]\\
k^{(LL)}&=&\tilde{k}^{(LL)}=h^{(LL)}-\frac{3}{(x-1)}+\frac{3x\log x}{(x-1)^2}\\
k^{(LL\pr)}&=&h^{(LL\pr)}-\frac{3x}{(x-1)}+\frac{3x\log x}{(x-1)^2}\\
k^{(LL\pr\pr)}&=&h^{(LL\pr\pr)}-\frac{3}{(x-1)}+\frac{3x\log x}{(x-1)^2}\\
k^{(LRt)}&=&\tilde{k}^{(LRt)}=\tilde{k}^{(LRt\pr)}=h^{(LRt)}+\frac{3}{x-1}-\frac{3x\log x}{(x-1)^2}
\end{eqnarray}

\end{document}